\documentclass[a4paper]{jpconf}
\usepackage{graphicx}
\RequirePackage{lineno}
\RequirePackage{latexsym}
\RequirePackage{xspace}
\RequirePackage[colorlinks,citecolor=blue,urlcolor=blue,linkcolor=blue]{hyperref}
\usepackage{amsmath,amsfonts,bm}
\usepackage{wasysym}
\usepackage{upgreek}
\usepackage{lmodern}
\usepackage{anyfontsize}

\newcommand{\onbb}{$0\nu\beta\beta$\xspace}
\newcommand{\nnbb}{$2\nu\beta\beta$\xspace}
\newcommand{\thalf}{$T_{1/2}^{0\nu}$}

\newcommand{\Qbb}{$Q_{\beta\beta}$\xspace}

\newcommand{\Mo}{$^{100}$Mo\xspace}
\newcommand{\lmo}{Li$_2$MoO$_4$\xspace}
\newcommand{\enrLMO}{Li$_{2}${}$^{100}$MoO$_4$\xspace}

\newcommand{\Tl}{$^{208}\mathrm{Tl}$\xspace}
\newcommand{\Bi}{$^{214}\mathrm{Bi}$\xspace}

\newcommand{\gabe}{$\gamma$\&$\beta$\xspace}

\newcommand{\ckky}{counts/(keV$\times$kg$\times$yr)\xspace}

\begin{document}
\title{First data from the CUPID-Mo neutrinoless double beta decay experiment}
\author{B. Schmidt on behalf of the CUPID-Mo Collaboration \footnote{Collaboration list: http://cupid-mo.mit.edu/collaboration}}

\address{Nuclear Science Division, Lawrence Berkeley National Laboratory, Berkeley, California 94720, USA }
\ead{beschmidt@lbl.gov}

\begin{abstract}

\noindent The CUPID-Mo experiment is searching for neutrinoless double beta decay in \Mo, evaluating the technology of cryogenic scintillating \enrLMO detectors for CUPID (CUORE Upgrade with Particle ID). 
CUPID-Mo detectors feature background suppression using a dual-readout scheme with \lmo crystals complemented by Ge bolometers for light detection.  The detection of  both heat and scintillation light signals allows the efficient discrimination of  $\alpha$ from \gabe events.
In this proceedings, we discuss results from the first 2 months of data taking in spring 2019. In addition to an excellent bolometric performance of 6.7\,keV (FWHM)  at 2615\,keV and an $\alpha$ separation of better than 99.9\% for all detectors, we report on bulk radiopurity for Th and U. Finally, we interpret the accumulated physics data in terms of a limit of \thalf $\,> 3\times10^{23}$\,yr for \Mo and discuss the sensitivity of CUPID-Mo until the expected end of physics data taking in early 2020.

\end{abstract}

\section{Introduction}

In 1937 Ettore Majorana published a seminal paper on a ``Symmetrical theory of electrons and positrons'' \cite{Majorana:1937}. The theory predicts new lepton number violating processes and has been investigated for its possibilty of producing the matter-antimatter asymmetry in the Universe through leptogenesis \cite{Fukugita:1986}.
It also predicts that neutrinos can be their own antiparticles giving rise to the hypothetical decay process of neutrinoless double beta decay (\onbb). In this decay  the (anti-) neutrino is reabsorbed in the second vertex and  only two electrons and the nucleus remain in the final state. Hence, the full decay energy can be easily detected \cite{DellOro:2016tmg}.
Modern \onbb searches are an extremely sensitive probe of the light Majorana neutrino exchange mechanism in particular and of new lepton number violating physics in general.   
Bolometric detectors, such as CUORE, are among the leading present searches \cite{ Gando:2016,Albert:2018,Aalseth:2018,Alduino:2018,Agostini:2018}. These detectors feature one of the best energy resolutions in the field, very high signal efficiency and they excel in the range of target isotopes that can be studied \cite{Giuliani:2018}.
A ton-scale cryogenic search has been demonstrated \cite{Alduino:2018} and an additional background suppression by the discrimination of $\alpha$ versus \gabe events has been established in both R\&D \cite{Armengaud:2017} and in medium scale demonstrator experiments \cite{Azzolini:2019nmi, CUPIDMo:2019}. 
The \enrLMO detectors operated in CUPID-Mo have been selected for the future upgrade for CUORE since they feature an  excellent  performance  and radiopurity \cite{Armengaud:2017, Poda:2017a}. In the following we will present performance results as well as an evaluation of the physics data from the first 2 months of the physics campaign from April to June 2019.

\section{Performance of scintillating \lmo bolometers}
The CUPID-Mo experiment employs twenty $\sim$210$\,$g \lmo crystals enriched to $\sim  97\,\%$ in \Mo. The detectors modules are arranged into 5 towers of 4 modules, where each module comprises a \enrLMO crystal at the top and an adjacent thin Ge light detector below. 

This detector arrangement is housed inside of the low radioactivity infrastructure of the EDELWEISS-III experiment \cite{Armengaud:2017b, Hehn:2016} in the underground laboratory of Modane (France). The setup and the detector performance were characterized in an initial commissioning campaign in March 2019 and have been described in detail in \cite{CUPIDMo:2019}. All detectors were found to be operational and only a single detector module with abnormal noise condition was discarded from the analysis. Of the remaining 19 detectors all achieved performances, compatible with better than 99.9\,\% $\alpha$ separation at equally high \gabe acceptance. 
In the present analysis we use the same 19 detectors and present results from an extended period of data taking between April and June 2019, about 5 times longer than the comissioning. 
The data processing is based on the Diana processing tools developed by the CUORE and CUPID-0 collaborations \cite{Alduino:2016, Azzolini2018b}.

\begin{figure}[h]
\begin{minipage}[t]{0.45 \textwidth}
\includegraphics[ trim=0 0 0 30, clip, width=0.95 \textwidth]{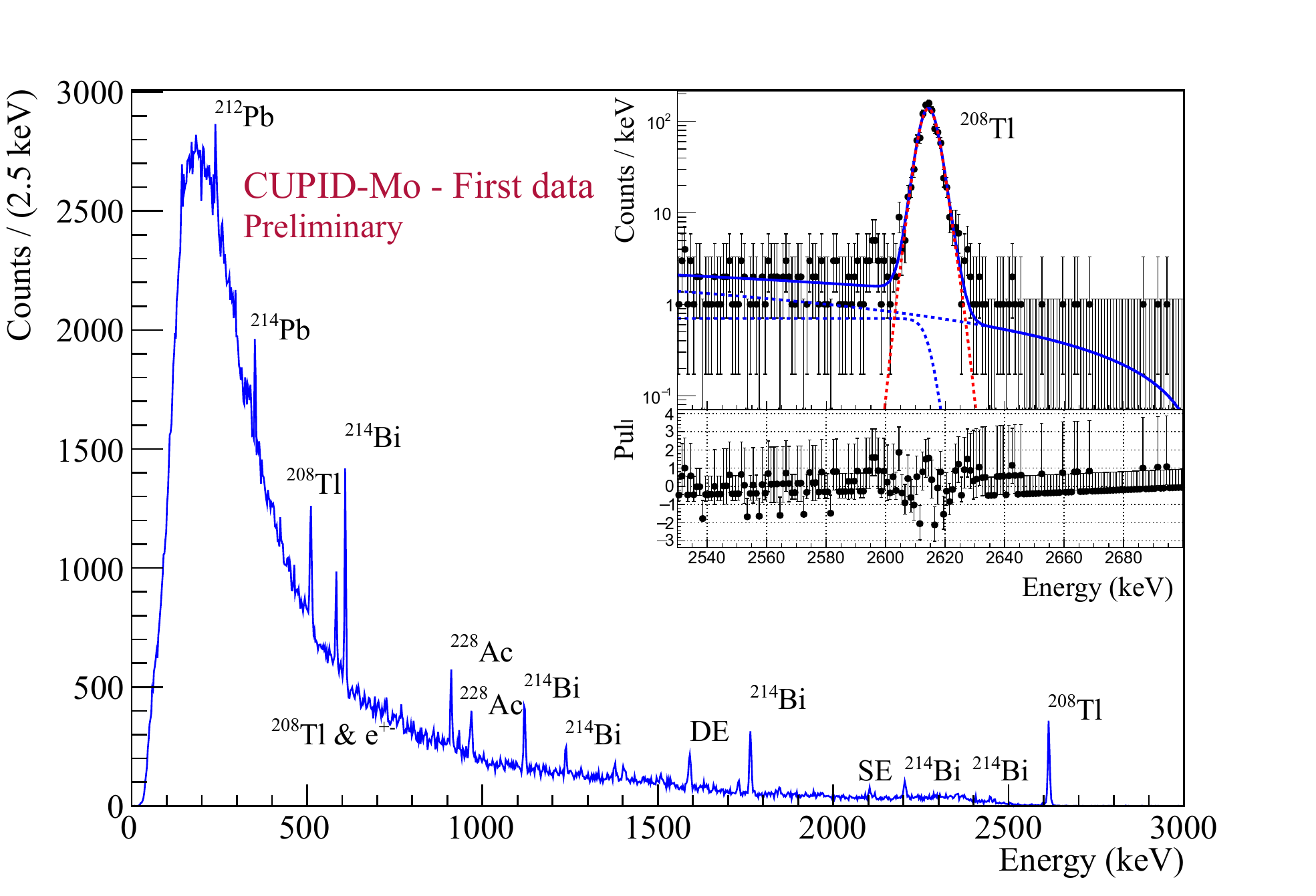}
\caption{\label{fig:CalSpec} Summed calibration data of 19 detectors accumulated over 10 days of U/Th calibration. Significant peaks  have been labeled. The inset shows the result of a simultaneous UEML fit of the 2615 keV \Tl peak. }
\end{minipage}\hspace{0.05 \textwidth}%
\begin{minipage}[t]{0.45 \textwidth}
\includegraphics[ trim=0 0 0 30, clip, width=0.95 \textwidth]{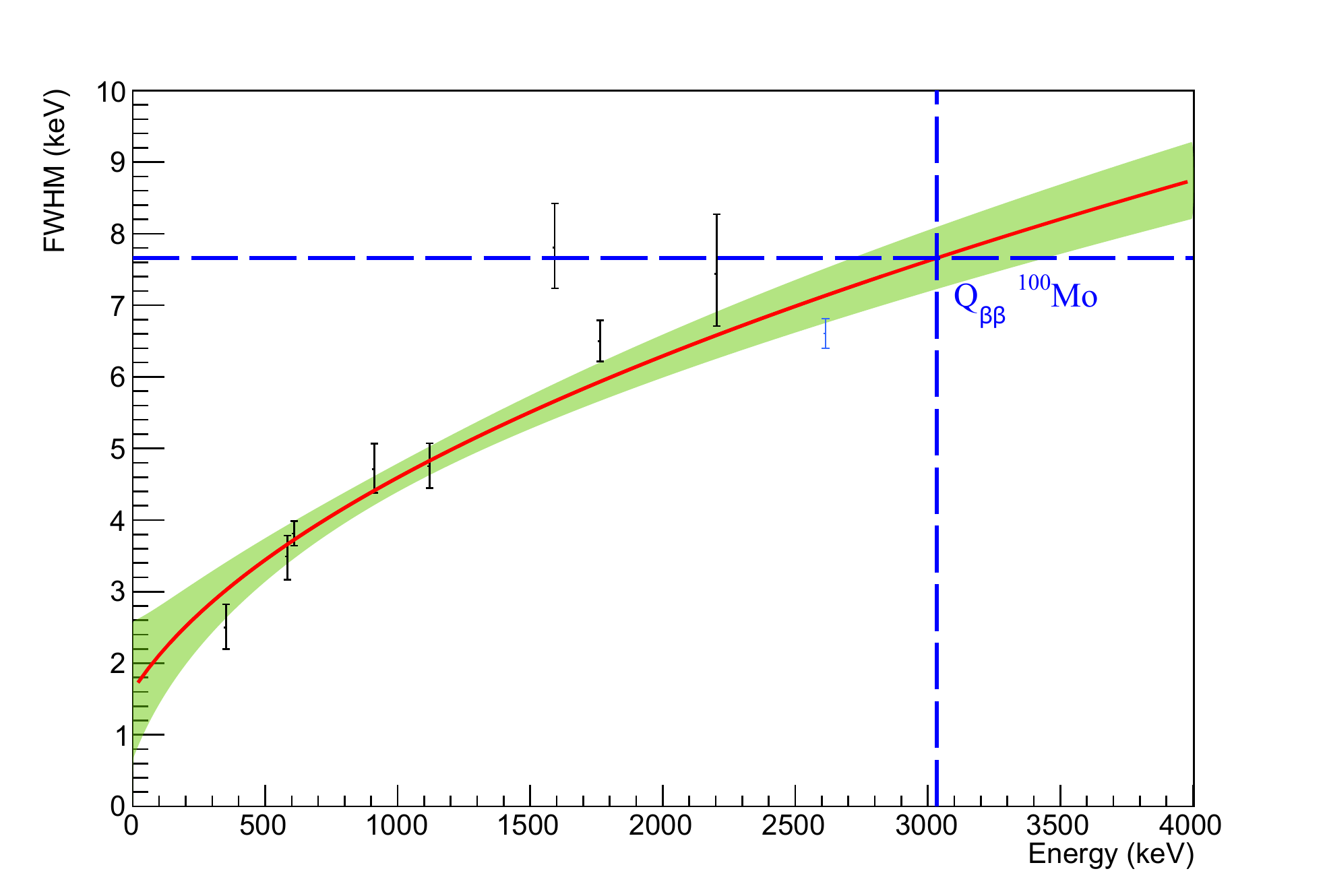}
\caption{\label{fig:ResolutionScaling} Energy dependence of the resolution of the CUPID-Mo detectors, modeled with $\sigma(E) = \sqrt{\sigma_0^2 + \left(\sigma_E*\sqrt{E}\right)^2}$. 
The fit result is drawn in red, with the 1\,$\sigma$ uncertainty band in green. }
\end{minipage} 
\end{figure}

In addition to the previous linear calibration \cite{CUPIDMo:2019} we perform a first non-linear calibration of the \enrLMO detectors. Using the most prominent lines from \Tl (2615\,keV) and \Bi (609\,keV, 1120\,keV, 1764\,keV) we fit a 2nd order polynomial through their respective mean positions and obtain the summed calibration spectrum, shown in Fig. \ref{fig:CalSpec}. 
We perform a simultaneous unbinned extended maximum likelihood (UEML) fit across the 19 detectors with individual amplitudes and resolutions for the \Tl peak and extract a resolution of 6.7 keV FWHM (harm. mean) at 2615 keV. We further model the spectrum with a common step function that describes multi-Compton events in terms of the ratio of the continuum to the \Tl peak height and we include a common linear component to approximate \nnbb events and remaining background from pile-up.
We use significant $\gamma$-peaks of the summed calibration spectrum to obtain resolution estimates for various energies in order to extrapolate the energy resolution at 3034\,keV (\Mo \Qbb). We evaluate both a linear fit as well as an extrapolation adding the resolution from a fixed noise component and a statistical term that scales with $\sqrt{E}$ in quadrature. In Fig. \ref{fig:ResolutionScaling} the preferred fit is drawn, which predicts a resolution of $(7.7\pm0.4)\,$keV (FWHM) at \Qbb.

\section{First physics data}
\begin{figure}[h]
\begin{minipage}{0.49 \textwidth}
\includegraphics[width=0.99 \textwidth]{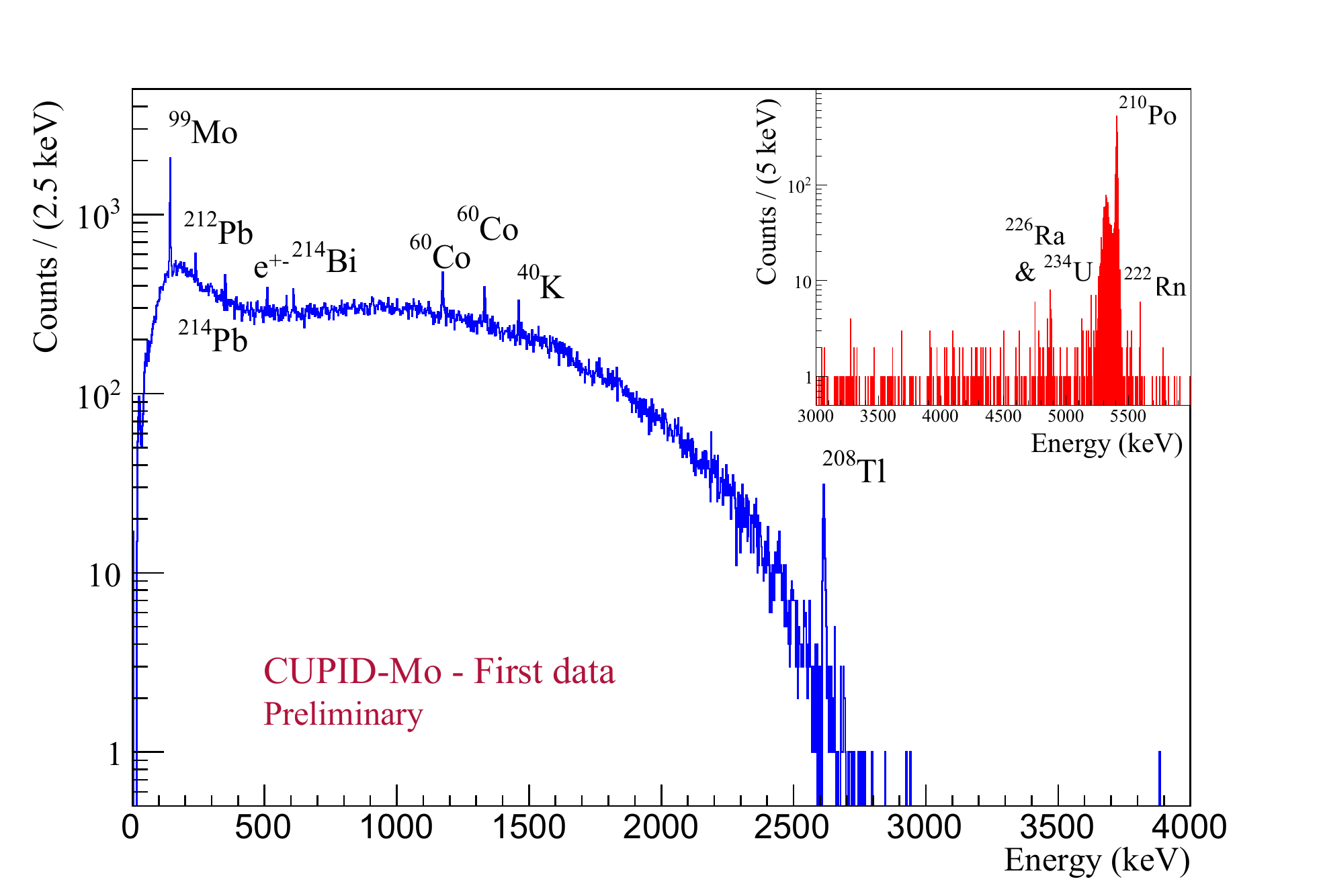}
\caption{\label{fig:BgSpec} Summed background spectrum for 19 detectors and 0.5\,kg $\times$ yr of exposure after analysis cuts. The inset shows the alpha region (3\,MeV -- 6\,MeV) with basic data quality cuts only.  Significant peaks have been labeled. }
\end{minipage}\hspace{0.05 \textwidth}%
\begin{minipage}{0.45 \textwidth}
\includegraphics[width=0.99 \textwidth]{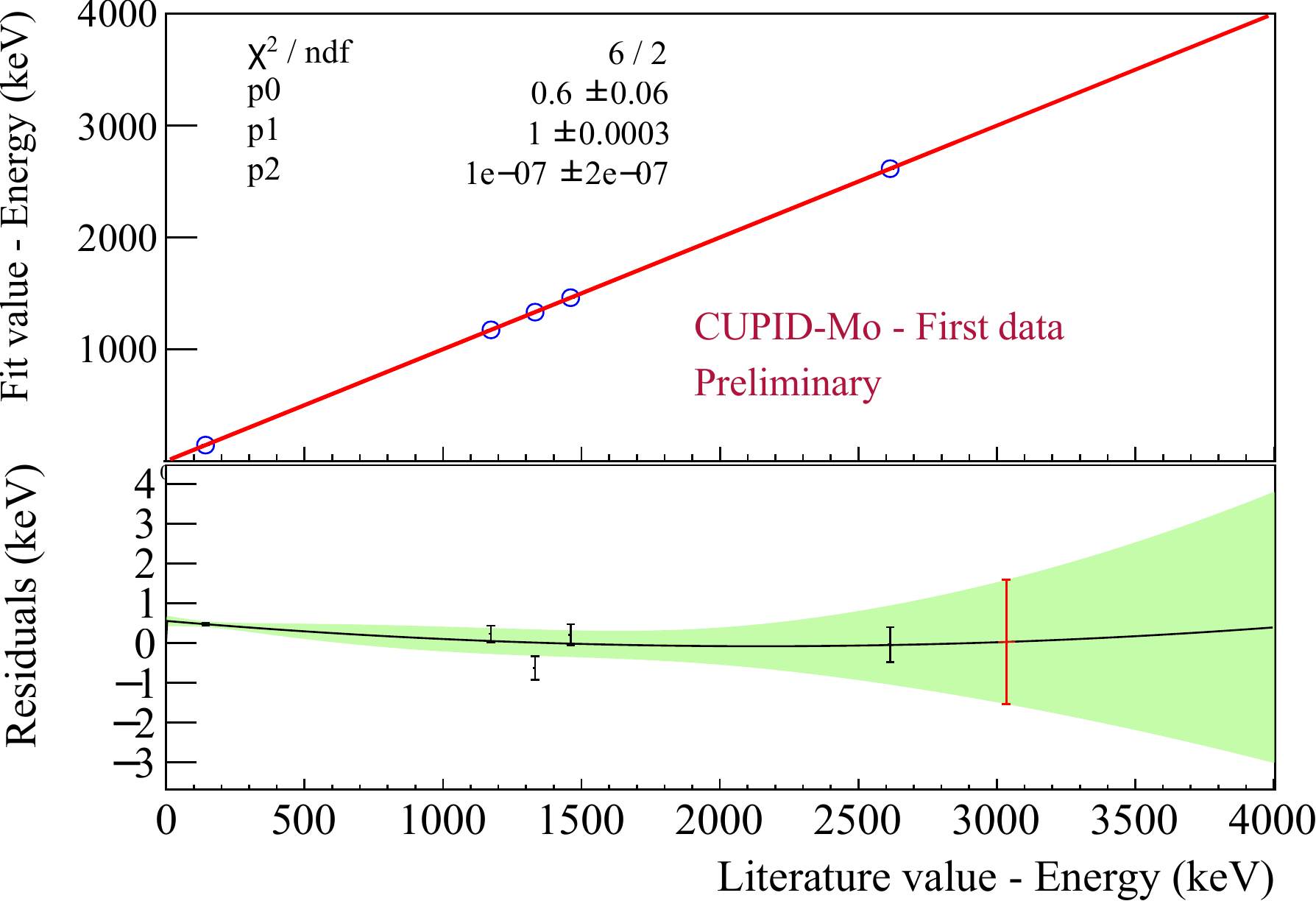}
\caption{\label{fig:BgEnergyScale}Energy linearity of the background data with residuals from a 2nd order polynomial fit with respect to the true value in the lower inset. The 1\,$\sigma$ confidence band (green) and the extrapolated bias and uncertainty at \Qbb (red) are shown.}
\end{minipage} 
\end{figure}

The accumulated exposure in physics data amounts to $0.5\,$kg$\times$yr after removal of unstable cryogenic periods and periods with abnormal noise conditions (4$\%$). 
In addition to the previous analysis procedure of the commissioning data, we defined four pulse shape analysis cuts on normalized pulse shape variables (rise-time, decay-time, trigger - optimum filter position, baseline slope). These cuts have been set at $\pm 7$ median absolute deviations (MAD), however they presently dominate our efficiency at a level of $\epsilon=(81.1 \pm 0.5) \%$. We thus plan to further study our pulse shape normalization procedure and thoroughly optimize our cuts for a future blind \onbb analysis. The resulting spectrum for \gabe events, dominated by the \nnbb decay shape
is shown in Fig. \ref{fig:BgSpec}.

In order to prepare for the \onbb analysis we evaluate the accuracy of our energy scale on five significant peaks in the background data (Fig. \ref{fig:BgEnergyScale}), fitting a 2nd order polynomial in estimated versus true peak position. The quadratic term is compatible with 0 and the estimated bias at \Qbb is $\Delta E = (0 \pm 1.6)\,$keV.
 Relaxing all pulse shape and pile-up cuts we perform a preliminary investigation of the bulk contamination of the \enrLMO crystals counting events in a $\pm$ 15 keV window from the alpha decays in the U-series  and the Th-series. We obtain results of $^{238}$U - $(0.5 \pm 0.2)\,\mu$Bq/kg, sum of $^{234}$U+$^{226}$Ra - $(1.4 \pm 0.3)\,\mu$Bq/kg , $^{230}$Th - $(0.3 \pm 0.14)\,\mu$Bq/kg, $^{222}$Rn - $(0.6 \pm 0.2)\,\mu$Bq/kg, $^{218}$Po - $(0.5 \pm 0.2)\,\mu$Bq/kg for the U-series and similarly $^{232}$Th - $(0.3 \pm 0.14)\,\mu$Bq/kg, $^{228}$Th - $(0.4 \pm 0.2)\,\mu$Bq/kg, $^{224}$Ra - $(0.25 \pm 0.12)\,\mu$Bq/kg, $^{212}$Bi $(0.3 \pm 0.2)\,\mu$Bq/kg for the Th-series.

\section{Outlook}
\label{sec:Conclusion}
\begin{figure}[htbp]
  \centering
  \includegraphics[width=0.55\textwidth]{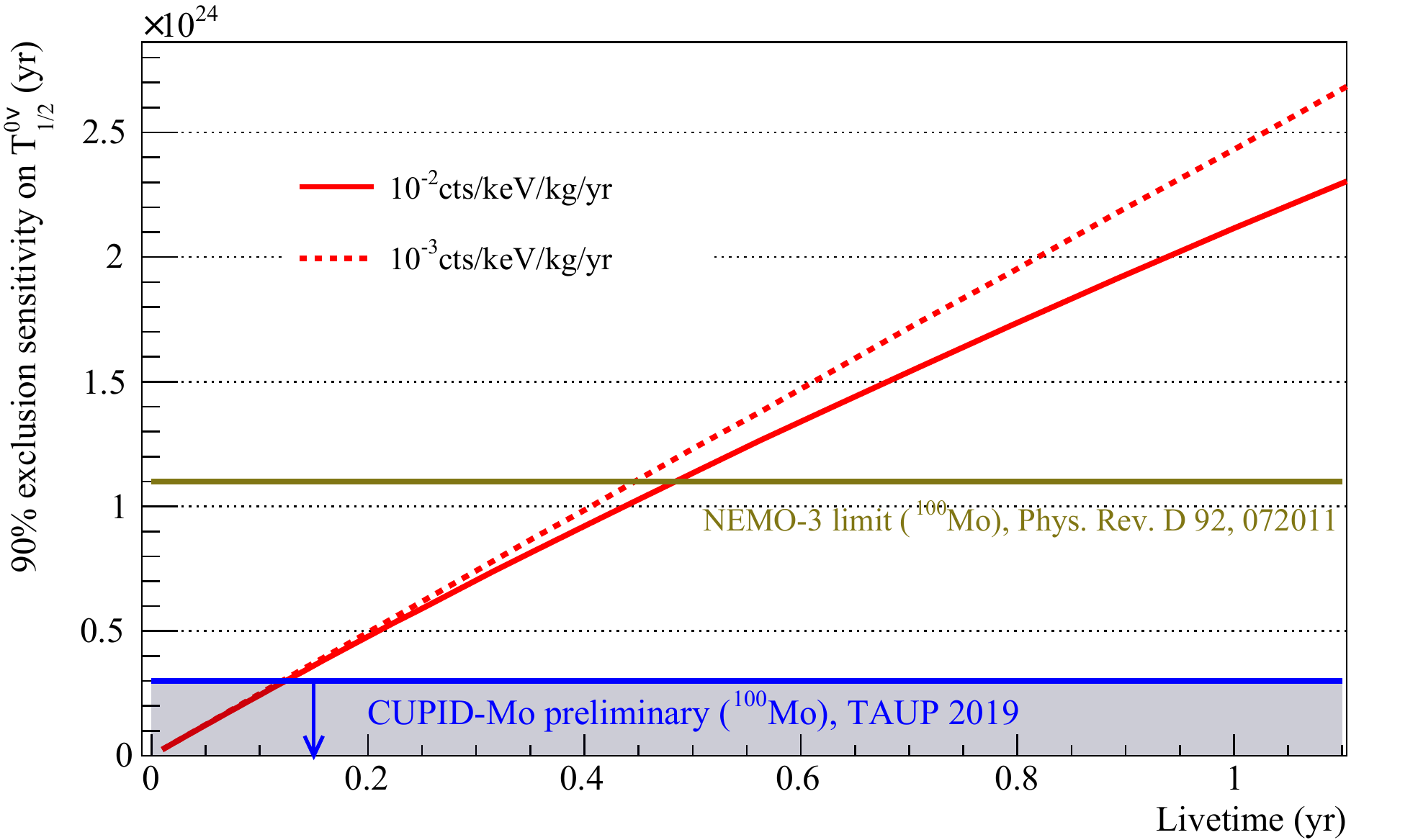}
  \caption{Bayesian exclusion sensitivity at 90\%\,C.I. for 5\,keV resolution (FWHM) at \Qbb and different background levels. Both the present limit after 2 months of data taking and the NEMO-3 limit after 5 years of data taking with $\sim$7\,kg of \Mo are shown.}
  \label{fig:sensitivity}
\end{figure}

Based on these first two months of physics data we can confirm the high degree of reproducibility and ease of operation of the detectors. Thanks to the very low background with a single event between 3 MeV and 4 MeV and no event close to \Qbb of \Mo at $3034\,$keV we can translate these first results into a limit on \thalf\ using a one sided 90\% Poisson confidence bound of 2.3 events. We fold in the systematic uncertainty from the efficiency evaluation following 
Cousins and Highland \cite{Cousins1992} and obtain a limit of \thalf $> 3 \times 10^{23}\,$ yr (see Fig. \ref{fig:sensitivity}).
We compare this result to our projected sensitivity calculated for a 5\,keV energy resolution (FWHM),  75\% \onbb containment efficiency, a $\sim$ 90\% analysis efficiency and for the background index of either $10^{-2}$ \ckky or $10^{-3}$ \ckky. Since the present exposure is in the background free regime for both hypothesis, there is no impact of the slightly worse energy resolution and we fall within 10\% of the expected exclusion limit. We note that this limit is a factor 4 worse than the leading limit of NEMO-3 for \Mo of \thalf $> 1.1 \times 10^{24}\,$ yr \cite{Arnold:2015}, achieved after 5 years of data taking with $\sim$7 instead of $\sim$2 kg of \Mo. 
Based on these very encouraging results from a preliminary analysis we started working on the development of a detailed background model 
and to optimize the procedure for a blind \onbb analysis with more than half a year (2 kg $\times$ yr) of physics exposure in early 2020. 
\newline \\ \vspace{0.3cm}
\small
\noindent \textbf{Acknowledgements:} \newline
We gratefully acknowledge the help of the technical staff of the Laboratoire Souterrain de Modane and other participating laboratories as well as our supporting funding agencies. For the full Acknowledgement please see {\small \url{http://cupid-mo.mit.edu/collaboration}}. 

\section*{References}
\bibliographystyle{iopart-num}
\bibliography{Bibliography}

\end{document}